\documentclass[conference]{IEEEtran}

\usepackage{graphicx}
\usepackage{caption}
\usepackage{subcaption}
\usepackage{color}
\ifCLASSINFOpdf
\else
\fi

\usepackage{amsmath}
\usepackage{algorithmic,algorithm}
\allowdisplaybreaks
\begin{document}

\title{Maximum Throughput Scheduling for Multi-connectivity in Millimeter-Wave Networks}

\author{\IEEEauthorblockN{Cristian Tatino\IEEEauthorrefmark{1}\IEEEauthorrefmark{2}, Ilaria Malanchini\IEEEauthorrefmark{2}, Nikolaos Pappas\IEEEauthorrefmark{1}, Di Yuan\IEEEauthorrefmark{1}}
\IEEEauthorblockA{\IEEEauthorrefmark{1}Department of Science and Technology, Link\"{o}ping University, Sweden\\
               Email: \{cristian.tatino, nikolaos.pappas, di.yuan,\}@liu.se}      
               \IEEEauthorblockA{\IEEEauthorrefmark{2}Nokia Bell Labs, Stuttgart, Germany\\
               Email: ilaria.malanchini@nokia-bell-labs.com}
}

\maketitle
\IEEEpeerreviewmaketitle
\begin{abstract}
Multi-connectivity is emerging as promising solution to provide reliable communications and seamless connectivity at the millimeter-wave frequency range. Due to the obstacles that cause frequent interruptions at such high frequency range, connectivity to multiple cells can drastically increase the network performance in terms of throughput and reliability by coordination among the network elements. In this paper, we propose an algorithm for the link scheduling optimization that maximizes the network throughput for multi-connectivity in millimeter-wave cellular networks. The considered approach exploits a centralized architecture, fast link switching, proactive context preparation and data forwarding between millimeter-wave access points and the users. The proposed algorithm is able to numerically approach the global optimum and to quantify the potential gain of multi-connectivity in millimeter-wave cellular networks.
\end{abstract}

\section{Introduction}
\label{sec:Intro}
The fifth generation (5G) of mobile communications is characterized by ambitious requirements to be fulfilled in terms of broadband access, connection density and data rate, e.g., $50$ Mbps everywhere and an amount of connections in the order of thousands per km\textsuperscript{2} as reported in~\cite{5G}. Millimeter wave (mm-wave) communications technology is one of the promising solutions to tackle these challenges. Mm-wave increases the available spectrum resources by exploiting under-utilized frequency bands between $30$ to $300$ GHz. On the other hand, communication in these frequency bands is subject to more challenging propagation conditions than the lower frequency bands, especially in terms of free space path loss and penetration loss. This latter may lead to frequent transmission interruptions when there is no line-of-sight (LOS). Multi-connectivity (MC) represents a possible solution to minimize the number of communication interruptions, by allowing the users (UEs) to establish and maintain connections with multiple cells/access points at the same time. 

The benefits of having multiple links available for a single UE in a mm-wave wireless network are discussed in several works, e.g., \cite{Div} and~\cite{MultiBlo}. These show how the MC can improve the mm-wave performance in terms of denial of service, outage and drop session probabilities when multiple mm-wave access points (mmAPs) can simultaneously transmit to the UE. 
The work in~\cite{Danish} proposes several potential centralized architectures for MC in mm-wave communications. In particular, these are based on a common framework for mm-wave access supported by low band frequencies. The authors in~\cite{Danish} qualitatively analyze the different architectural approaches in terms of required messages and overhead. 
Moreover, in~\cite{MultiZo}, the authors propose a new alignment procedure, which can represent a bottleneck in a mm-wave MC scenario. The proposed method can reduce the duration of this phase when the UEs sustain connections with multiple mmAPs. 
The vast majority of the works dealing with MC assume a centralized architecture with accurate synchronization between the network elements, which is also a fundamental requirement for the coordinated multipoint (CoMP) joint transmission method. In particular, the work in~\cite{COMP1} proposes a centralized radio-over-fiber architecture for CoMP in mm-wave small cell networks. The analysis and the experimental results show increased performance in reception quality and probability of LOS.

In this work, we consider a two-layer outdoor cellular network consisting of 5G low band base stations (5G-LBs) and mmAPs connected to a network controller (NC). Moreover, by exploiting the synchronization between the mmAPs, we assume that a UE can establish and maintain multiple connections and possibly receive data from several mmAPs at the same time, similar to a CoMP joint transmission approach.

\subsection{Contributions}
\label{sec:Contributions}
In a multi-connectivity scenario, one of the most relevant aspects is the link scheduling problem. In this work, we propose a novel link scheduling approach for network throughput maximization, and quantify the potential gain of MC and CoMP for mm-wave cellular networks. We formulate the optimal link scheduling problem as a binary integer program (BIP) and then, we propose an algorithm for solving the problem, which is able to numerically approach the global optimum. We exploit a fast link switching mechanism based on the assumption that each link can be in one of four different states. These states determine the time needed to the link to establish a connection and start the transmission. 

Finally, by using predicted information about the channel quality over a certain time window, the NC optimizes the link scheduling and prepares in advance the UEs and the mmAPs for the communication. We evaluate the performance of the proposed scheduling algorithm with simulations, and we show the advantages with respect to the single connectivity (SC) approach in terms of throughput and number of interruptions. The rest of the paper is organized as follows: in Section~\ref{sec:Ass} we describe the system model and the assumptions. In Section~\ref{sec:Prob} and in Section~\ref{sec:Opt} we formulate the optimization problem and we present the solution algorithm for the MC approach. Finally, Section~\ref{sec:Perf} illustrates the results and performance comparison and Section~\ref{sec:Conc} concludes the paper.

\section{System Model and Assumptions}
\label{sec:Ass}
As in~\cite{Danish}, we consider an MC approach for which each UE is associated to a 5G-LB and possibly to several mmAPs, where, the control and the data planes are split and using the 5G-LB and the mmAPs, respectively. The two layers are connected to an NC in a centralized architecture that enables fast link switching, mmAPs data buffer synchronization, data forwarding, and possibly joint transmissions. In this latter case, several mm-wave connections transmit to the UE the same data, combined then at the receiver~\cite{COMP1}. 

Furthermore, we assume that the mmAPs are perfectly synchronized both in frequency and time and the signals arrive to the UE in-phase. The beamforming gain of downlink transmissions at the receiver depends on how the UE and the mmAPs perform beam alignment. For this purpose, both the mmAP and the UE have several sub-arrays and radio frequency (RF) chains such that each of them can perform the beam alignment with a mmAP, independently.
Moreover, we consider that the beams are narrow enough so that the interference becomes negligible, as shown in~\cite{Int}, and we assume that the signal is completely blocked in NLOS conditions and the resulting link throughput is zero.

We assume that the NC has the perfect knowledge of the channel quality condition for each UE over a certain upcoming time window in term of both LOS/NLOS states and achievable throughput. In a real scenario these information can obtained by applying several channel quality estimation and prediction techniques. Since, in this work, we are mainly interested to show the potential gain of MC, we do not consider a specific prediction technique, but the NC exploits this information and proactive link preparation, for optimizing link scheduling. 

\subsection{Fast Link Switching}
\label{sec:state}
As mentioned above, we consider a fast link switching mechanism, based on~\cite{Danish}, which exploits different link states. In particular, each link state defines which operations the mmAP, the UE and NC are performing, i.e., UE context transferring, data buffer exchanging and transmission.
In this work, we use the term UE context, for both 5G-LB and mmAPs, to refer to the block of information required to maintain the UE connection, as defined in~\cite{3GPP2}. The UE context is established with a 5G-LB or mmAPs at the first access of a UE as well as when the handover procedure is completed. Hereafter, the following link states are introduced: active (A), hot stand-by (H), cold stand-by (C), and inactive (I).
\begin{itemize}
\item The link is in A when the mmAP is transmitting data, received by the NC, to the UE. 
\item A link is in H when the UE, the mmAP and the NC have completed the UE context preparation. The mmAP is receiving UE data from the NC, but it is not transmitting them, since the beam alignment is not yet performed. 
\item A link is in C when the mmAP is not receiving data from the NC. The mmAP is performing only the operations needed to obtain the UE context information and the buffering data, which are not received yet. 
\item A link is in I if it is in none of the above states.
\end{itemize}
The mmAP consumes power, backhaul capacity, and radio resources (time-frequency slots) in both A and H states. The purpose of introducing four different states is to define the transition of a link from I to A. Namely, each link in I should go through C and H before becoming A. By allowing a link to be prepared in advance, we can reduce the time needed to activate the link when a communication interruption occurs. The transition between H and A can occur only in LOS (since it requires pilot transmission~\cite{Init}), while, the transition from C to H can be performed in NLOS since the control messages between the NC and UE are transmitted by the 5G-LB. We refer to the amount of time needed to transit between states as transition time, which depends on the performed operations.

In summary, given the assumptions and the architecture described above, we aim to solve the optimal link scheduling problem in an MC mm-wave cellular network such that, given a certain time window for which the channel quality is known, the network throughput is maximized. The link scheduling problem is subject to the constraints on the transition time between the different link states and the mmAP total power budget. In the following sections, we formulate the optimization problem and present a solution algorithm.

\section{Multi-connectivity Problem Formulation}
\label{sec:Prob}
In this section, we formulate the optimization problem introduced in Section~\ref{sec:state}. Namely, given a set of mmAPs~$\mathcal{M}$, a set of UEs~$\mathcal{U}$, and a set of time slots $\mathcal{K}$, with cardinalities $M$, $U$ and $K$ respectively, the NC decides, for each time slot, which state to assign to each link in order to  maximize the network throughput (i.e., the sum of all the UEs' throughput). 
The relation between the UE's throughput and the corresponding active links is non-linear. To deal with non-linearity, we first introduce the concept of \textit{configuration}, which can be then used to formulate the optimization problem as a linear BIP.
A configuration refers to a subset of links whose elements are active and in LOS. Since the channel condition may change over time, for slot $k$ we define the set of all the possible configurations $\mathcal{S}_{k}$ and denote its cardinality by $S_{k}$. Each configuration $s \in \mathcal{S}_{k}$ defines the achieved network throughput.

Now, given the sets of configurations $\mathcal{S}_{k}$, the task of the NC is to choose one among the given configurations (at each time slot) in order to maximize the network throughput. As mentioned above, we assume that at each time slot and for each mmAP-UE pair, the NC knows the link status, i.e., whether or not it is in LOS. Therefore, we define the binary parameter $l^{ijk}$, which is equal to $1$ if link $ij$ between mmAP $i$ and UE $j$ is in LOS in time slot $k$ and $0$ otherwise. 
Moreover, since a link can be in several states, as described in Section~\ref{sec:state}, we define the mutually exclusive binary variables $y^{ijk}_{a}$, $y^{ijk}_{h}$, and $y^{ijk}_{c}$. They represent the state of link $ij$ in time slot $k$, and they are equal to $1$ when the link states are A, H and C, respectively. When the variables associated to a link in a time slot are all equal to zero, the link is inactive.

Furthermore, we define the variable $z^{sk}$, with $s \in \mathcal{S}_{k}$ and $k \in \mathcal{K}$, which is set to $1$ if configuration $s$ is used in time slot $k$ and $0$ otherwise. 
Since each configuration defines a subset of active links, we introduce the parameter $\alpha^{ijsk}$, which is equal to $1$ if link $ij$ is active in configuration $s$ of time slot $k$. 
The throughput provided by a configuration $s$ to UE $j$ in time slot $k$ is given by parameter $r^{jsk}$. This value is known to the NC, because the throughput is fully determined in each configuration. Now, we can formulate the problem as follows: 
\begin{subequations}
\begin{align}
        P1:&\max_{y^{ijk}_{a},y^{ijk}_{h},y^{ijk}_{c},z^{sk}} \sum_{k=1}^{K} \sum_{s=1}^{S_{k}}   \sum_{j=1}^{U} r^{jsk} z^{sk} \label{opt}\\ 
        \text{s.t.}&\sum_{j=1}^{U} (P_{a}y^{ijk}_{a}+P_{h}y^{ijk}_{h})\le P_{tot} \quad \forall i\in\mathcal{M},k\in\mathcal{K} \label{Pot}\\
				&y^{ijk}_{a}+y^{ijk}_{h}+y^{ijk}_{c} \le 1 \quad \forall i\in\mathcal{M},j\in\mathcal{U},k\in\mathcal{K} \label{mut}\\
				&\sum_{s=1}^{S_{k}}z^{sk} \le 1 \quad \forall k\in\mathcal{K} \label{confc}\\
				&y^{ijk}_{a}= \sum_{s=1}^{S_{k}} \alpha^{ijsk}z^{sk} \quad \forall i\in\mathcal{M},j\in\mathcal{U},k\in\mathcal{K} \label{active}\\
				&l^{ijn}y^{ijn}_{h}+y^{ijk-1}_{a}\ge y^{ijk}_{a} \label{tran_a}\\ &\qquad \forall i\in\mathcal{M},j\in\mathcal{U},k\in\mathcal{K},n\in\{k-t_{ha},\ldots,k-1\} \nonumber \\
				&y^{ijn}_{c}+y^{ijk-1}_{h}+y^{ijk-1}_{a}\ge y^{ijk}_{h} \label{tran_h} \\ &\qquad \forall i\in\mathcal{M},j\in\mathcal{U},k\in\mathcal{K},n\in\{k-t_{ch},\ldots,k-1\} \nonumber\\
				&y^{ijk}_{a}, y^{ijk}_{h}, y^{ijk}_{c}, z^{sk}\in\{0,1\} \label{type} \\ &\qquad \forall i\in\mathcal{M},j\in\mathcal{U},k\in\mathcal{K}, s\in\mathcal{S}_{k} \nonumber.
\end{align}
\end{subequations}
The objective function, given by~\eqref{opt}, represents the sum of all the UEs' throughput associated to the selected configurations. 
The UE's throughput is computed by the well-known Shannon's formula for the AWGN channel capacity. According to the assumptions in Section~\ref{sec:Ass}, the throughput provided by the $s$-th configuration at the $k$-th time slot to the $j$-th UE can be computed as in~\cite{COMP1} :
\begin{equation}
\label{eq:rate}
r^{jsk}=B\log_{2}(1+\sum^{M}_{i=1}\alpha^{ijsk}SNR^{ijk}),
\end{equation}
where, $SNR^{ijk}$ is the signal-to-noise ratio for the link between the $i$-th mmAP and the $j$-th UE at the $k$-th time slot and $B$ denotes the system bandwidth.
Constraints~\eqref{Pot} limit the total transmit power per mmAP to be less than or equal to the power budget, where $P_{a}$ and $P_{h}$ are parameters that represent the power consumed by a mmAP when a link is A or H, respectively. Inequalities~\eqref{mut} impose each link to be in maximum one state. Constraints~\eqref{confc} force the selection of at most one configuration per time slot, for which the corresponding active links are set by equalities~\eqref{active}.  

Constraints \eqref{tran_a} and \eqref{tran_h} take into account the transition times from state H to A and from C to H, respectively. In particular, link $ij$ can be active in time slot $k$ if either it is A in time slot $(k$-$1)$ or it is H for the previous $t_{ha}$ time slots consecutively. This comes from the alignment phase, in which the mmAP and the UE need to be in LOS. Similarly, a link can be in the state H if either it is in the state A or H in time slot $(k$-$1)$ or it has been in the state C for $t_{ch}$ time slots. 

The optimization problem P1 is not convex since it is a BIP. Moreover, the cardinality of $\mathcal{S}_{k}$, that is at most $2^{MU}$, grows exponentially with the number of UEs and mmAPs. However, most of the configurations are not relevant for constructing the optimal solution. Therefore, we follow an approach similar to the one considered in~\cite{Di} that is presented in the next section.


\section{Multi-connectivity Optimization Algorithm}
\label{sec:Opt}
In this section, we present an algorithm to solve the problem P1 formulated in Section \ref{sec:Prob} and deal with the exponential growth of configurations. The algorithm is based on a column generation approach. Specifically, we consider the continuous relaxation of P1, which is a linear program, and decompose it into a master problem (P2) and a pricing problem (P3). For sake of space we do not explicitly show the formulation of P2, but it can be expressed as P1, with the differences that (a) the constraint \eqref{type} is replaced with $0 \le y^{ijk}_{a}, y^{ijk}_{h}, y^{ijk}_{c}, z^{sk} \le1$ (continuos relaxation) and (b) we consider only a subset of configurations $\breve{\mathcal{S}_{k}}$. By solving P3, we generate new configurations and associated network throughput that progressively improve the solution of P2. 

More precisely, the configurations represent columns in P2 and adding a new configuration is equivalent to adding a new column. We generate a new subset of configurations per each time slot $k$ since the channel conditions, hence the network throughput associated with a configuration, may change over the time slots. The solution of P2 is used by the pricing problem to determine whether a new configuration can improve network throughput or not. The resulting algorithm is depicted in Algorithm~\ref{alg:Algo} and described in details in Section~\ref{sec:Algo}.

\subsection{Pricing Problem}
\label{sec:Conf}
In this section, we define the pricing problem for the column (i.e., configuration) generation. Before proceeding with the formulation, we introduce the concept of local enumeration as in~\cite{Di}. This is needed in order to decrease the complexity of the pricing problem, since for each UE there are $2^{M}$ possible configurations of active links, and considering all of them is not scalable. The local enumeration concept is based on the idea that, in each time slot, typically the number of mmAPs, that significantly contribute to the UE's throughput, is small with respect to $M$.
\begin{figure}[tb]
	\centering
	\includegraphics[width=8cm]{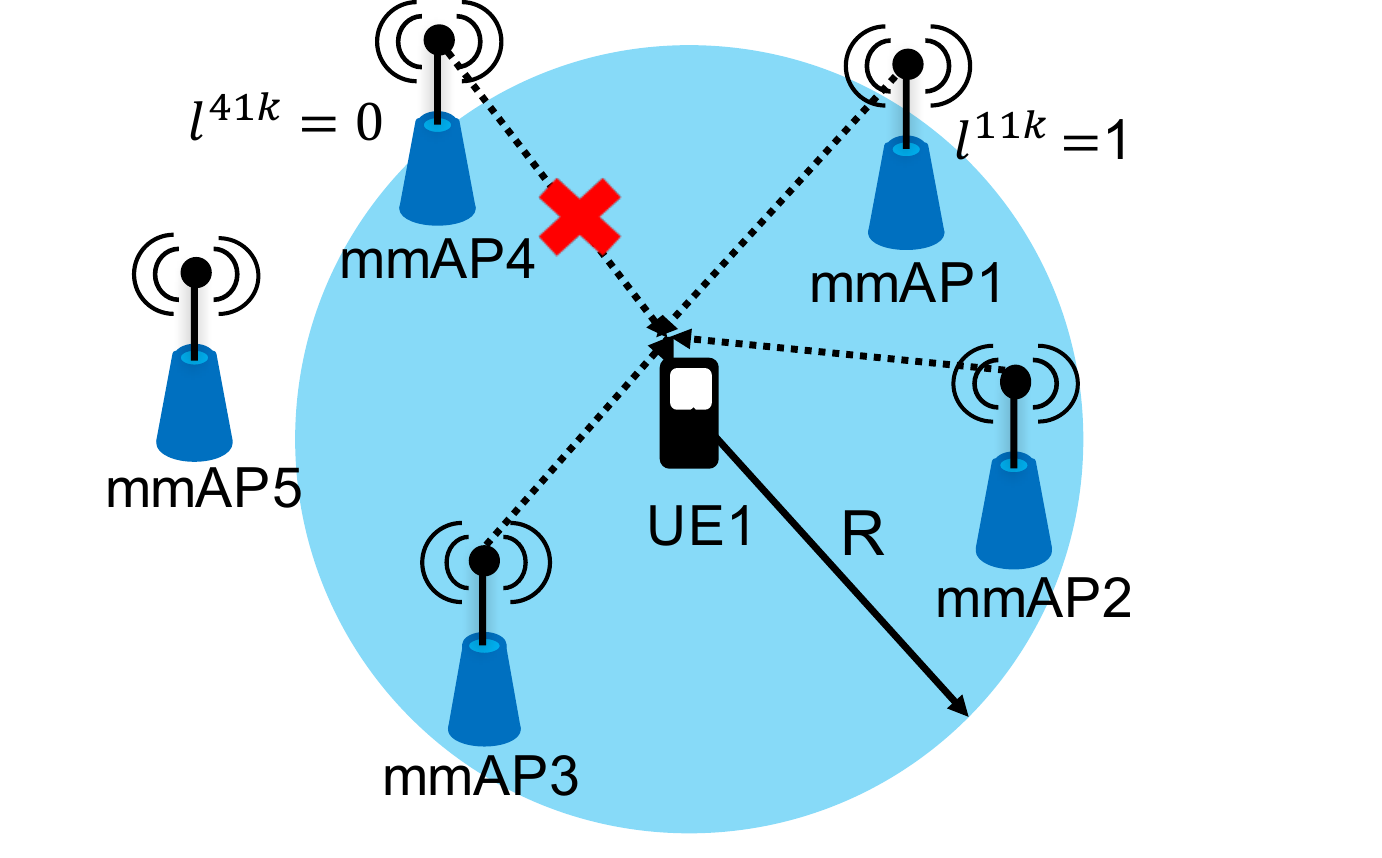}
	\caption[]{Possible scenario for the local enumeration. In this case only mmAPs $1$, $2$ and $3$ are considered for UE $1$ since mmAP $4$ is in NLOS and mmAP $5$ is outside the range R, thus, only mmAPs $1$, $2$, and $3$ are included in the set $\mathcal{Q}^{1k}$. Therefore, we have that $E_{1}=2^3$, and the possible combinations are represented by the following sets: $\mathcal{V}^{11k}=\{\}$, $\mathcal{V}^{12k}=\{1\}$, $\mathcal{V}^{13k}=\{2\}$, $\dots$, $\mathcal{V}^{18k}=\{1,2,3\}$.
}
	 \label{fig:Loc}
\end{figure}
Thus, we define the set $\mathcal{Q}^{jk}$, which contains all the mmAPs that are in LOS and at a maximum distance $R$ with respect to UE $j$ in time slot $k$. Then, we consider all the possible combinations of the active mmAPs included in the set $\mathcal{Q}^{jk}$, whose number is $E_{j}$. Each combination is represented by a set $\mathcal{V}^{jek}$, which contains the active links for UE $j$ in time slot $k$ for combination $e \in [1 \ldots E_{j}]$. Each set $\mathcal{V}^{jek}$ determines a UE's throughput $r^{jek}$. Fig.~\ref{fig:Loc} shows an example of local enumeration for UE $1$ with some possible combinations of active links.

Now, in order to formulate the pricing problem, we introduce the binary variables $\alpha^{ij}$ and $v^{je}$. The former is set to 1 if the link between the mmAP $i$ and the UE $j$ is A, and zero otherwise, and it is then used to set the corresponding parameter $\alpha^{ijsk}$ of the master problem (for the considered time slot $k$ and generated configuration $s$). The latter variable, $v^{je}$ determines which set $\mathcal{V}^{jek}$ is selected among the $E_{j}$ scenarios for the active links of UE $j$. Thus, we can formulate the pricing problem for time slot $k$ as follows:
\begin{subequations}
\begin{align}
        P3:&\max_{\alpha^{ij},v^{je}} \sum_{j=1}^{U}r^{j} - \sum_{i=1}^{M}\sum_{j=1}^{U} \lambda^{ij}\alpha^{ij}  \label{subOb}\\ 
        \text{s.t.}&\sum_{j=1}^{U} P_{a}\alpha^{ij} \le P_{tot} \quad \forall i\in\mathcal{M} \label{subPot}\\
				&\sum_{e=1}^{E_{j}}v^{je} = 1 \quad \forall j\in\mathcal{U} \label{subConf}\\
				&\sum_{e:h\in \mathcal{V}^{je}}v^{je}  \le \alpha^{hj} \quad \forall j\in\mathcal{U}, h \in \mathcal{Q}^j \label{subAct}\\
				&r^{j}= \sum_{e=1}^{E_{j}}v^{je}r^{je} \quad  \forall j\in\mathcal{U}. \label{subRate}\\
				&\alpha^{ij},v^{je}\in\{0,1\} \quad \forall i\in\mathcal{M},j\in\mathcal{U}, e \in [1 \ldots E_{j}] \label{typeP}.
\end{align}
\end{subequations}
Expression~\eqref{subOb} represents the objective function, where $\lambda^{ij}$ are the dual variables associated with constraints \eqref{active} of P2. The dual variables $\pi^{ij}$, associated with constraints \eqref{confc} of P2, are not explicitly written out, because they are additive constants in the objective function.

Constraints \eqref{subPot} are related to the total transmit power budget for the mmAPs that is considered also here in order to generate only configurations that are feasible for P2. Equalities \eqref{subConf} impose that only one combinations of active links is selected per time slot and per UE. The constraints \eqref{subAct} identify which links must be active according to the selected combinations and state the consistency between the $v$-variables and $\alpha$-variables. The equalities \eqref{subRate} define the UEs' throughput $r^{j}$ associated with the selected combinations.

\subsection{Algorithm}
\label{sec:Algo}
Finally, the optimal scheduling algorithm is presented in Algorithm 1. At each iteration, Algorithm~\ref{alg:Algo} solves the problem P2 and obtains the dual variables $\lambda^{ij}$ and $\pi^{ij}$, whose values are used to compose the objective function of the pricing problem \eqref{subOb}. 
Thus, for each time slot, we solve the problem P3 and include the resulting configuration and associated UEs' throughput to the master problem if the reduced cost is positive. The algorithm concludes when none of the remaining configurations, not included in the subsets $\breve{\mathcal{S}_{k}}$, has positive reduced cost. In this case we solve the binary integer version of P2.

\begin{algorithm}
\caption{}
\label{alg:Algo}
\begin{algorithmic}[1]
  \scriptsize
  \STATE{Construct P2 with a subset of configurations $\breve{S_{k}}$}
  \REPEAT {
  	\FOR{$k\in K$}
		\FOR{$s\in S_{k} \backslash \breve{S_{k}}$}
			\STATE{Solve P2}
			\IF{$\max_{s\in S_{k} \backslash \breve{S_{k}}} \sum_{j=1}^{U}r^{jsk}-\sum_{i=1}^{M}\sum_{j=1}^{U} \lambda^{ijsk}\alpha^{ijsk} -\pi^{ij}> 0$}
				\STATE{Add the corresponding configuration to $\breve{S_{k}}$}
			\ENDIF 
		\ENDFOR
  	\ENDFOR
	 } \UNTIL{$\max_{k \in K}\max_{s\in S_{k} \backslash \breve{S_{k}}} \sum_{j=1}^{U}r^{jsk}-\sum_{i=1}^{M}\sum_{j=1}^{U} \lambda^{ijsk}\alpha^{ijsk} \leq 0$}
	 \STATE{Solve the binary integer version of P2}
\end{algorithmic}
\end{algorithm}

\section{Performance Evaluation}
\label{sec:Perf}
\begin{figure}[tb]
	\centering
	\includegraphics[width=8cm]{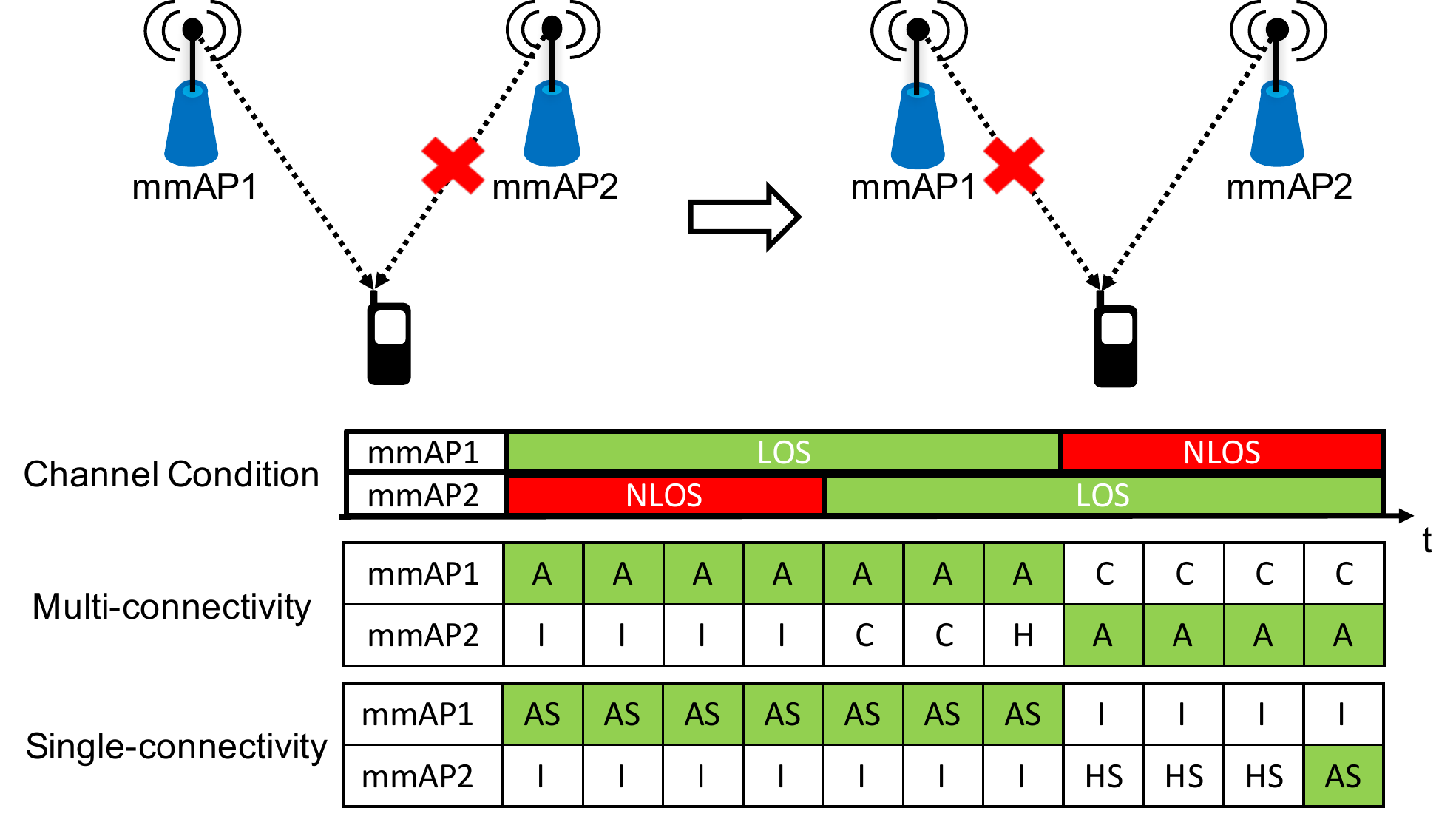}
	\caption[]{This figure shows the difference between the link switching of the MC and SC. In the figure, $t_{s}$ is equal to the sum of $t_{ha}$ and $t_{ch}$, which in turn are set to $1$ and $2$ time slots, respectively. For the SC case we define the states AS and HS, which correspond to a link that is active or performing the handover, respectively.}
	\label{fig:FigTra}
\end{figure}

In this section, we compare the maximum network throughput achieved by Algorithm~\ref{alg:Algo} (``MC + CoMP") and the solution of the SC approach. For sake of space we do not present the formulation of this latter, which can be written as a linear BIP.  More precisely, in the SC case, a UE can be connected only with one mmAP per time slot, which means that the UE can either receive data from or performing the handover with a mmAP. The handover duration is represented by the parameter $t_{s}$. In order to make a fair comparison between the MC and the SC cases, also for this latter we assume perfect channel quality prediction. In summary, for the SC case we have that (a) link switching can only start in LOS (there is no support of 5G-LBs) and (b) the UE cannot be connected to the previous mmAP while it is switching to a new one, as shown in Fig.~\ref{fig:FigTra}. 

Moreover, we compare these solutions with the network throughput reached by the MC case without CoMP (``MC no CoMP"), in which, differently from the SC case, the UEs can keep the connections with both the 5G-LB and possibly multiple mmAPs. However, for the MC without CoMP case, each UE can have at maximum one active link per time slot. In addition, we report the solution of the master problem P2, once the column generations is ended (``MC + CoMP Bound"), which represents an upper bound of the global optimum of P1.

\subsection{Simulation Setup}
\label{sec:SimEnv}
We consider an area of $250 \times 250$ m\textsuperscript{2}, where the mmAPs and the UEs are randomly distributed following a uniform distribution. The UEs move following a uniform rectilinear motion with a speed of $3$ km/h and a constant direction randomly chosen. Similar to \cite{Blo}, we model the interruption inter-arrivals as independent exponential random variables with parameter $\lambda$, while the duration of an interruption is characterized by a uniform distribution between $[400,1000]$ milliseconds (ms).

In order to compute the throughput associated to a link, we express the SNR in decibel (dB) as follows: $SNR=P_{a}+G_{a}+G_{u}-PL_{UMi-LOS}-P_{N}$, where $G_{a}$ and $G_{u}$ are the beamforming gains of the mmAP and the UE respectively. The term $PL_{UMi-LOS}$ is the path loss, which follows the model for urban micro cells (UMi) in LOS outdoor street canyon environment described in~\cite{3GPP}. This depends on the height of the mmAP $h_{BS}$, the height of the UE $h_{ut}$, the carrier frequency $f_{c}$ and the distance between the transmitter and the receiver. $P_{N}$ represents the noise power.

\begin{table}[!t]
	\renewcommand{\arraystretch}{1.3}
	\caption{Simulation parameters.}
	\label{par}
	\centering
	\begin{tabular}{| l | l || l | l || l | l |}
		\hline
		$P_{tot}$ & 30 dBm & $P_{a}$ & 24 dBm & $P_{h}$ & 24 dBm\\ \hline
		$G_{a}$ & 15 dBi & 	$G_{u}$ & 10 dBi & $h_{BS}$ & 10 m\\ \hline
		$h_{ut}$ & 1.5 m &	$f_{c}$ & 30 GHz &	$P_{N}$ & -85 dBm\\ \hline
		$B$ & 1 GHz &	$K$ & 20 & $t_{ha}$ & 1\\ \hline
		$t_{ch}$ & 2 &	$t_{s}$ & 3 & $R$ & 360 m\\ \hline
	\end{tabular}
\end{table}
\begin{figure}[tb]
	\centering
	\includegraphics[width=1\columnwidth]{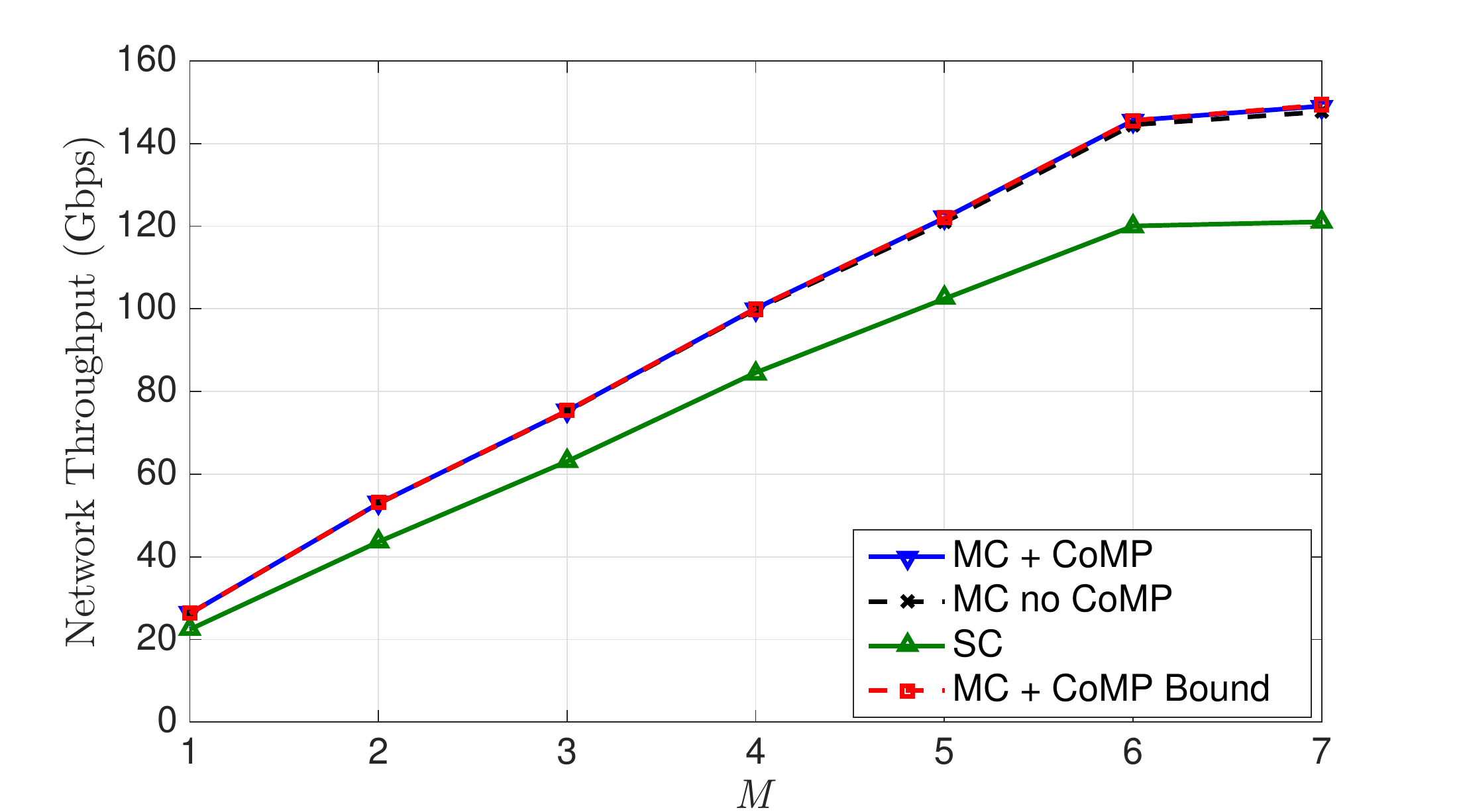}
	\caption[]{Network throughput when varying the number of mmAPs $M$, for $U$ = $20$, $\lambda^{-1}$ = $250$~ms.
	}
	\label{fig:M}
\end{figure}
In order to obtain a comprehensive performance evaluation, hereafter, we vary several parameters, such as $M$, $P_{tot}$, and $\lambda$. Unless specified otherwise, other parameters are fixed and shown in Table~\ref{par}, where $K$, $t_{ha}$, $t_{ch}$, and $t_{s}$ are expressed in the number of time slots. More precisely, for $t_{ha}$, we refer to the values described in~\cite{MultiZo} for digital beamforming at the mmAPs and analog beamforming at the UEs, with $32$ beams at the mmAP and $8$ at the UE. Since $t_{ha}$ is the smallest time unit in scheduling, we set the duration of a single time slot equal to $t_{ha}$ and we assume that the channel condition does not change in a time slot. Furthermore, we assume $R$ to be large enough for the considered area, so that all mmAPs can potentially transmit to all UEs.


\subsection{Results}
\label{sec:Res}
\begin{figure}[tb]
	\centering
	\includegraphics[width=1\columnwidth]{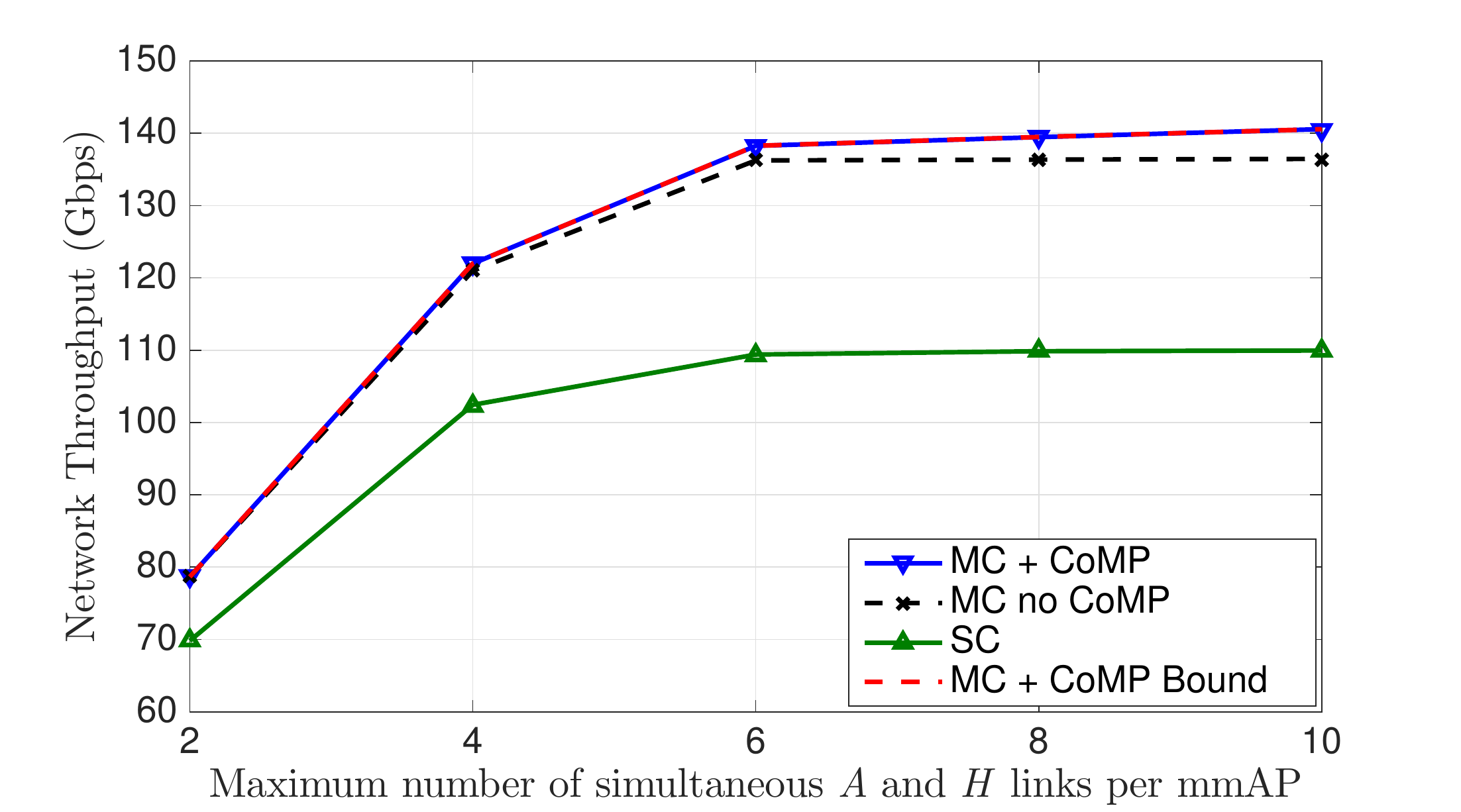}
	\caption[]{Network throughput when varying the total power budget expressed in terms of maximum number of simultaneous A and H links, for $M=5$ and $U=30$ and $\lambda^{-1}$ = $250$~ms.  
	}
	\label{fig:P}
\end{figure}

\begin{figure}[tb]
	\centering
	\includegraphics[width=1\columnwidth]{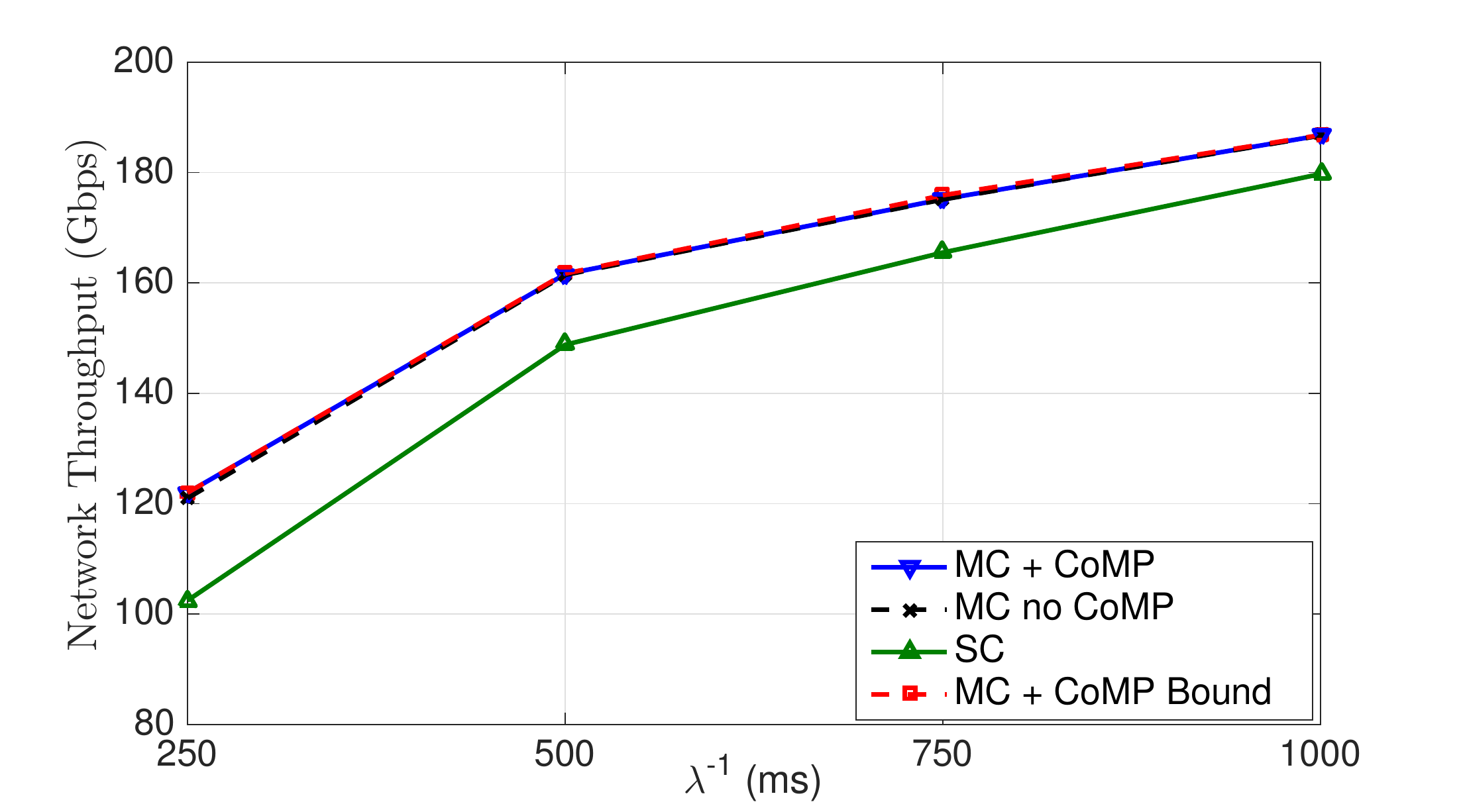}
	\caption[]{Network throughput when varying the interruption parameter $\lambda$, for $M=5$ and $U=30$.  
	}
	\label{fig:L}
\end{figure}

In Figs.~\ref{fig:M}-\ref{fig:L}, we compare the network throughput resulting from the MC with and without CoMP, and the SC cases, when varying $M$, $\lambda$ and $P_{tot}$. Note that, given the assumption $P_{a}$ = $P_{h}$, the power budget, $P_{tot}$, can be expressed in terms of maximum number of simultaneous A and H links per mmAP ($4$ for $P_{tot}=30$ dBm). We can observe that Algorithm~\ref{alg:Algo} almost overlaps with the upper bound, which means that, at least for the analyzed cases, the solution of Algorithm~\ref{alg:Algo} is able to numerically approach the global optimum.

Specifically, Fig.~\ref{fig:M} shows the network throughput when varying the number of mmAPs $M$. The number of UEs is $U$ = $20$ and $\lambda^{-1}$ = $250$~ms. We can observe that for both the MC and SC approaches the network throughput increases as the number of mmAPs grows. The MC cases (with and without CoMP) provide always a higher network throughput with respect to the SC approach, which is reflected also in the average number of interruptions per UE. Indeed, for $M = 5$ and $U = 20$ the number of time slots without any active links per UE is equal to $6$ and $9$ for the MC and SC cases, respectively. The gap between these latter at $M=1$ is due to the fact that even with only one mmAP, MC can rely on the support of the 5G-LB, which allows the links to be prepared in advance, even in NLOS. 

In Fig.~\ref{fig:M}, the difference between the solution of Algorithm~\ref{alg:Algo} and MC without CoMP is very small. For this reason, we can conclude that most of the gain of Algorithm~\ref{alg:Algo} with respect to the SC approach is given by the multiple connections, the fast link switching and the 5G-LB support. In this case, the improvement in terms of network throughput given by using CoMP transmissions is relatively small, which means that it is more efficient to serve more UEs with different links than to serve a lower amount of UEs by providing them multiple transmitting links. 
This is due to the relation between the active links and the UE's rate in the CoMP case, which is given by \eqref{eq:rate}. Moreover, given $P_{tot}$, $P_{a}$, and $P_{h}$, the maximum number of A and H links per time slot per mmAP is $4$ and the number of UEs is $U = 20$. Therefore, for most of the cases in Fig.~\ref{fig:M}, the number of LOS links is lower than the number of UEs and the NC does not take advantages from CoMP.  

Indeed, when we increase the maximum number of transmitting links per mmAP per time slot, i.e. $P_{tot}$, the difference between the MC without and with CoMP transmissions increases in favor of the latter. This is shown in Fig.~\ref{fig:P}.
Moreover, we can observe that, when the number of available links goes above a certain threshold, the slope of the curves decreases and, for any of the considered schemes, increasing $P_{tot}$ becomes less beneficial. This phenomenon (though less evident) appears also in Fig.~\ref{fig:M}, where the number of available links indirectly increases by increasing the number of mmAPs.
Finally, Fig.~\ref{fig:L} shows the comparison among the different approaches when varying the inter-arrival link interruptions rate $\lambda$ with a fixed number of UEs $U = 20$ and mmAPs $M = 5$. We observe that the gap between SC and MC cases decreases with $\lambda$ (i.e. when $\lambda^{-1}$ increases). As a matter of fact, when the interruptions become less frequent, the links are more stable and, therefore, the advantages of fast link switching are less evident.

\section{Conclusion}
\label{sec:Conc}
We have considered optimal link scheduling in an multi-connectivity mm-wave cellular network with the objective of maximizing the network throughput over a certain time window with constrained total power budget per mmAP. 
The proposed column generation approach in Algorithm~\ref{alg:Algo} leads to a solution that is very close or even equal to the global optimum. 
Numerical results show the potential gain of multi-connectivity in millimeter-wave cellular networks. The proposed solution results in significant improvement of the network throughput with respect to the SC case. The improvement is more profound when the number of mmAPs increases or the link interruptions are more frequent.

Extensions of this work include investigating the benefits of multi-connectivity when minimizing the number of interruptions and considering error-prone prediction of the channel quality information exploited by the network controller.
 

\section*{Acknowledgment}
The authors would like to thank Dr. Danish Aziz, Dr. Paolo Baracca, Dr. Lutz Ewe and Dr. Vangelis Angelakis for the insightful discussions. 
This work has received funding from the European Union's Horizon 2020 research and innovation programme under the Marie Sklodowska-Curie grant agreement No. 643002. Furthermore, this work has been partially supported by CENIIT.





%


\bibliographystyle{IEEEtran}
\bibliography{ref}

\end{document}